\documentstyle[12pt]{article}

\def\+{{+\!\!\!+}}

\def\d{\partial}

\def\m{\mu}
\def\n{\nu}
 
\def\g{\gamma} 
\def\G{\Gamma} 
 
\def\N{\nabla}

\def\de{\delta} 
\def\P{\Phi}

\def\r{\rho} 
\def\l{\lambda} 
 
\def\L{\Lambda} 
 
\def\s{\sigma} 
 
\def\F{\Psi} 
\def\e{\varepsilon}

\def\t{\tau} 
\def\k{\kappa}

\def\j{{\cal J}} 
\def\pmb#1{\setbox0=\hbox{#1}%
\kern.0em\copy0\kern-\wd0 
\kern-.04em\copy0\kern-\wd0 
\kern.08em\copy0\kern-\wd0 
\kern-.04em\raise.0433em\box0 }         
 
\def\frak#1#2{ \textstyle{#1\over #2}} 
\def\half{\frac{1}{2}}

\newcommand{\nc}{\newcommand} 
\nc{\beq}{\begin{equation}} 
\nc{\eeq}[1]{\label{#1}\end{equation}} 
\nc{\ber}{\begin{eqnarray}} 
\nc{\eer}[1]{\label{#1}\end{eqnarray}} 
\nc{\pek}[1]{\cite{#1}} 
\nc{\enr}[1]{(\ref{#1})} 
\nc{\kal}[1]{{\cal{#1}}} 
\nc{\dott}{\;\cdot\;} 
\newcommand{\Section}[1]{\section{#1} \setcounter{equation}{0}} 
 
\begin{document} 
\newcommand{\inv}[1]{{#1}^{-1}} 
\renewcommand{\theequation}{\thesection.\arabic{equation}} 
\newcommand{\be}{\begin{equation}} 
\newcommand{\ee}{\end{equation}} 
\newcommand{\bea}{\begin{eqnarray}} 
\newcommand{\eea}{\end{eqnarray}} 
\newcommand{\re}[1]{(\ref{#1})} 
\newcommand{\qv}{\quad ,} 
\newcommand{\qp}{\quad .} 
\begin{center} 
                                \hfill UUITP-02/04\\
                                \hfill HIP-2004-01/TH\\
                                \hfill   hep-th/0401100\\ 
\vskip .3in \noindent 
 
\vskip .1in 
 
{\large \bf{Generalized N=(2,2) Supersymmetric Non-linear Sigma Models}} 
\vskip .2in 
 
{\bf Ulf Lindstr\"om}$^a{} ^b$\footnote{e-mail address: ulf.lindstrom@teorfys.uu.se}

\vskip .15in 
 
\vskip .15in 
$^a${\em Department of Theoretical Physics, \\
Uppsala University,
Box 803, SE-751 08 Uppsala, Sweden }\\ 
\bigskip

and
\bigskip

$^b${\em HIP-Helsinki Institute of Physics\\
P.O. Box 64
FIN-00014 University of Helsinki, 
Suomi-Finland\\}

\bigskip
\end{center}
\abstract{We rewrite the ${\cal N}=(2,2)$  non-linear sigma model using auxiliary spinorial superfields defining  the model on ${\cal T}\oplus^ *{\cal T}$, where ${\cal T}$ is the tangent bundle of the target space ${\cal M}$.} This is motivated by possible connections to Hitchin's generalized complex structures. We find the general form of the second supersymmetry compatible with the known one for the original model.
\vfill
\eject

\Section{Introduction}

Supersymmetric non-linear sigma models have an interesting relation to complex geometry.
The geometry of the  target space ${\cal T}$ is restrictive, and depends on the amount of supersymmetry and on the dimension of the underlying space-time. E.g., in four dimensions one supersymmetry implies K\"ahler geometry in the target space and two supersymmetries imply Hyperk\"ahler geometry. In two dimensions the situation becomes particularly interesting due to the relation to string theory and to the richer target space geometry allowed. In the pioneering paper \cite{Gates:nk} it was shown that the relevant geometry for (2,2) supersymmetry is a bi-hermitean geometry involving two complex structures, and that this geometry under certain circumstances become a generalization of K\"ahler geometry. Similar results hold for, e.g., (4,4) supersymmetry. 

In recent mathematical literature, in connection with Calabi-Yau manifolds, a new geometric structure called generalized complex geometry has been studied \cite{nigel,marco}. Among other features, it involves  a generalized complex structure defined on ${\cal T}\oplus^ *{\cal T}$. There are indications that this kind of geometry should be relevant for ${\cal N}=(2,2)$ supersymmetry in two dimensions. To understand this relation, one should generalize the usual sigma model to one that is defined on ${\cal T}\oplus^ *{\cal T}$ and investigate its geometry. In this note we perform the first of these tasks.

To this end, we rewrite the sigma model introducing two auxiliary spinorial superfields $\F_\pm$, in $^*{\cal T}$ apart from the usual scalars $\P$ in ${\cal T}$. Integrating out these fields the usual $2D$ supersymmetric sigma model is recovered, keeping them we may ask under which conditions this system has ${\cal N}=(2,2)$ supersymmetry. A priori, this larger system, involving extra fields, could introduce a more general geometry while still retaining the same amount of supersymmetry. 

When investigating the symmetries of the model, we are guided by the fact that by going partially on shell (using the $\F$ equations) we recover the known model involving only $\P$'s. We thus make an ansatz for the second supersymmetry (involving all fields $\P$ and $\F$) as general as allowed by dimensional analysis and the fact that  on $\F$-shell it must reduce to the known second supersymmetry for $\P$. Further, as usual, the pure $\P$ formulation of the sigma model forces us also on $\P$-shell for the supersymmetry algebra to close, giving additional relations.
In this manner, using a $1.5$ order formalism for the $\F$ fields and requiring both invariance of the action and on-shell closure of the algebra and enforcing a certain discrete symmetry of the action, we determine part of the transformations. The undetermined part is then shown to correspond to ``field equation''-type symmetries and removed. 
The final form (\ref{ppsec}) of the transformations is completely determined in terms of the complex structures characterizing the original $\P$-model, the metric $G$ and the antisymmetric tensor field $B$.

Some background material on supersymmetric sigma models is collected in section 2, the actual model is introduced in section 3, the additional supersymmetry is derived in section 4 and further discussed in section 5. After a quick look at isometries duand duality in section 6 we give our conclusions in section 7.
\Section{N=(2,2) sigma models, generalities}
In this section we summarize some basic facts about supersymmetric nonlinear sigma models.

The ${\cal N}=(1,1)$ action for a supersymmetric non-linear sigma model in a background metric $G_{\m\n}$ and antisymmetric 
$B_{\m\n}$ field reads
\beq
    S= \int d^2\xi d^2\theta\,\, D_{+} \Phi^\mu D_{-} \Phi^\nu E_{\mu\nu}
    (\Phi),
    \eeq{goodaction}
where $D_\pm$ are the spinorial derivatives satisfying the supersymmetry algebra $D^2_+=i\d_\+$, $D^2_-=i\d_=$, and
 where the metric $G_{\m\n}=\frac{1}{2}E_{(\m\n)}$ and the
  torsion potential
  $B_{\m\n}=\frac{1}{2}E_{[\m\n]}$ \footnote{ We use (anti-) symmetrization
 without a combinatorial factor}.

  As first described in \cite{Gates:nk},
  the action (\ref{goodaction}) has ${\cal N}=(2,2)$ supersymmetry
  \footnote{The target-space geometry for models with less supersymmetry,
  e.g, $(2,1)$, is also very interesting, but will not be discussed here. See \cite{Abou-Zeid:1999em}.} i.e.,
  an additional non-manifest supersymmetry of the form
  \beq
  \de \P^\m =
  \e^+(D_+\P^\n){\j}_\n^{(+)\m}+\e^-(D_-\P^\n){\j}_\n^{(-)\m}~,
  \eeq{secondSUSY}
  provided that $\j^{(\pm)}$ are complex structures:  they square to minus
  one,
  \beq
  \j^{2(\pm)}=-{\b1}~,
  \eeq{Cstruc}
  and have vanishing Nijenhuis tensors\footnote{More general models with non-vanishing ${\cal N}$ have
  also been considered \cite{Delius:nc}};
  \beq
  {\cal N}_{\m\n}^{~(\pm)\kappa}\equiv\j_\m^{(\pm)\g}\d_{[\g}\j_{\n]}^{(\pm)\kappa}-(\m\leftrightarrow
  \n)=0~.
  \eeq{Nij}
  In addition, the metric has to be {\em bi-hermitean}, i.e., hermitean with respect to {\em both} complex
  structures
  \beq
  \j_\m^{(\pm)\g}G_{\g\r}\j_\n^{(\pm)\r}=G_{\m\n}~,
  \eeq{biherm}
  and the complex structures should be covariantly constant with respect to certain
  connections $\G^{(\pm)}$, respectively
  \beq
  \N_\m^{(\pm)}\j_\n^{(\pm)\g}=0~.
  \eeq{CovJ}
  These  connections are
  \beq
  \G_{\m\n}^{(\pm)\g}=\G_{\m\n}^{(0)\g}\pm T_{\m\n}^{~\g}~,
  \eeq{Connex}
  with $\G^{(0)}$ the Christoffel connection for the  metric $G$, and the
 torsion
  given by
  \beq
  T_{\m\n}^{~\g}=\frac 1 2 H_{\m\n\r}G^{\r\g}~.
  \eeq{tor}
  This  relates the complex structures to the field-strength for the $B$-field,
  \beq
  H_{\m\n\r}=\d_{[\m}B_{\n\r]}~,
  \eeq{H}
  which implies
 \beq
H_{\m\n\r}=\j_\m^{(+)\g}\j_\n^{(+)\kappa}\j_\r^{(+)\l}d\j^{(+)}_{\g\kappa\l}=-\j_\m^{(-)\g}\j_\n^{(-)\kappa}\j_\r^{(-)\l}d\j^{(-)}_{\g\kappa\l}~,
  \eeq{HJrel}
  where $d\j^{(\pm)}$ is the exterior derivative of the two forms with
 components
  $\j_{\m\n}^{(\pm)}=\j_\m^{(\pm)\g}G_{\g\n}$,)\footnote{For a recent discussion of the relevant geometry, see \cite{Lyakhovich:2002kc}}.

When the two complex structures commute,
$[\j^{(+)},\j^{(-)}]=0$, their product gives an almost product
structure, i. e., $\Pi^2=1$ where
$\Pi=\j^{(+)}\j^{(-)}$ \cite{Gates:nk}. While the
individual integrability of $\j^{(+)}$ and $\j^{(-)}$ is not
sufficient to guarantee integrability of $\Pi$, in
conjunction with \re{CovJ} it is \cite{Gauntlett:2003cy}. We may
then choose coordinates where $\Pi$ is diagonal. It is this case which is possible to formulate in terms of
chiral and twisted chiral $N=2$ superfields \cite{Gates:nk}. More general models may be constructed using (anti-)semichiral superfields \cite{Buscher:uw} as
coordinates, as discussed in \cite{Ivanov:ec}. (For $N=4$ the
geometric structure is even more restricted \cite{Gates:nk}, and
there are additional superfield coordinates available
\cite{Lindstrom:mw})

\Section{The alternative action}

For the action (\ref{goodaction}) above,
the (bosonic part of) the superfields $\Phi^\mu$ coordinatize the target space ${\cal M}$, and their derivatives lie in the 
tangent bundle ${\cal T_M}$ of ${\cal M}$. In this paper we want to consider an action given by
\beq
    S= -\int d^2\xi d^2\theta\,\, \left\{\F_{+\m} \F_{-\n} E^{\mu\nu}(\Phi)+i\F_{(+\m}D_{-)} \Phi^\m\right\}~,
\eeq{betteraction}
where $E^{\mu\nu}$ is the inverse of $E_{\mu\nu}$ and the (bosonic part of) spinorial superfields $\F_{+\m}$ lie in the co-tangent space ${{}^*\cal T_M}$.
 The action (\ref{betteraction}) may thus be considered as a sigma model on  
${\cal T_M}\oplus{{}^*\cal T_M}$ and is equivalent to (\ref{goodaction}), as is seen by eliminating $\F_{\pm\m}$ 
via their field equations. The purely bosonic part reads
\beq
    S= \int d^2\xi \left\{A_{\+\m} A_{=\n} E^{\mu\nu}(X)+A_{[\+\m}\d_{=]} X^\m\right\}~,
\eeq{betterboson}
where $A_{\+\m}$ and $A_{=\n}$ are vector fields, (the $\theta^\pm$ components of $\F_{\pm\m}$), and $X^\m$ is the lowest component of $\P^\m$. When $E^{\mu\nu}$ is purely antisymmetric this is a Poisson sigma model \cite{Schaller:1994es} and (\ref{betteraction}) is its supersymmetric version. In what follows we invesigate the conditions for (\ref{betteraction}) to have  additional supersymmetry (${\cal N}=(2,2)$).

\Section{Transformations}

In this section we first make an ansatz for the second supersymmetry transformation and then use the requirements of closure of the algebra and invariance of the action to determine the coefficient functions. We shall need to discriminate between using all or only some of the field equations of (\ref{betteraction}). Henceforth the case  when both the $\P$ and the $\F$ equations are satsisfied will be refered to as being {\em on-shell} the case when only the $\F$ equations are satisfied will be refered to as being {\em on $\F$-shell}.

Since we shall only require the algebra to close on-shell, we use our knowledge of the transformations of the action (\ref{goodaction}) to which our action reduces on $\F$-shell. A further simplification will result from the discrete symmetry
\bea
&&\F'_{+\m}=-\F_{+\m}+ 2iD_+ \P^\n E_{\n\m}\cr
&&\F'_{-\m}=-\F_{-\m}- 2iD_- \P^\n E_{\m\n}\label{fredef}
\eea
which leaves the action (\ref{betteraction}) invariant.

By dimensional arguments, the most general transformations read

\ber
&&\delta \P^\m=\e^+D_+\P^\n J_\n^{~\m}-i\e^+\F_{+\r} E^{\r\n}I_\n^{~\m}\cr 
&&\delta \F_{+\m}=i\e^+\d_{\+}\P^\n M_{\n\m}+\e^+D_+\F_{+\n} K^\n_{~\m}\cr
&&~~+\e^+\F_{+\r}\F_{+\n}N^{\r\n}_{~~\m}+\e^+D_{+}\P^\r D_{+}\P^\n P_{\r\n\m}
+\e^+D_+\P^\r \F_{+\n}Q_{\r\m}^{~~\n}\cr
&&\delta \F_{-\m}=\e^+D_+\F_{-\n}R^\n_{~\m}+\e^+D_-\F_{+\n}S^\n_{~\m}\cr
&& ~~+\e^+D_+D_-\P^\n T_{\n\m}+\e^+\F_{+\r}D_-\P^\n U^\r_{~\n\m}
+\e^+D_+\P^\r \F_{-\n} V^{~~\n}_{\r\m}\cr
&&~~+\e^+D_+\P^\r D_-\P^\n X_{\r\n\m}+\e^+\F_{+\r}\F_{-\n}Y^{\r\n}_{~~\m}\label{tfans}
\eer{ansatz}
and similarily for $\e^-$.
On $\F$-shell we find from the $\F$-field equations that
\ber
&&\F_{+\m}=i D_+\P^\n E_{\n\m},\qquad D_+\F_{+\m}= -\d_\+\P^\n E_{\n\m}-iD_+\P^\r D_+\P^\n E_{\r\m\n},\cr 
&& D_-\F_{+\m}=-i D_+D_-\P^\n E_{\n\m}-iD_+\P^\r D_-\P^\n E_{\r\m\n}\cr
&&\F_{-\m}=-i D_-\P^\n E_{\m\n}\qquad D_-\F_{-\m}= \d_=\P^\n E_{\m\n}+iD_-\P^\r D_+\P^\n E_{\m\r\n},\cr
&& D_+\F_{-\m}=-i D_+D_-\P^\n E_{\m\n}+iD_-\P^\r D_+\P^\n E_{\m\r\n}
\eer{fieldeqns}

When requiring closure of the algebra, we will need the on $\F$-shell transformations which read

\bea
&&\delta \P^\m=\e^+D_+\P^\n (J_\n^{~\m}+I_\n^{~\m})\\ \nonumber
&&\delta \F_{+\m}=\e^+\d_{\+}\P^\n \L^{~\r}_\n E_{\r\m}+\e^+D_{+}\P^\r D_{+}\P^\n \tilde P_{\r\n\m}\\ \nonumber
&&\delta \F_{-\m}=\e^+D_+D_-\P^\s\tilde T_{~\s\m}+\e^+D_+\P^\r D_-\P^\n \tilde X_{\r\n\m}
\eea

where
\bea
&&\tilde P_{\r\n\m}\equiv P_{\r\n\m}+iE_{\n\s} Q_{\r\m}^{~~\s}-E_{\r\s}E_{\n\k}N^{\s\k}_{~~\m}-iE_{\r\s\n}K^\s_{~\m}~,\cr
&&\tilde T_{\s\m}\equiv -iE_{\n\s}R^\n_{~\m}-iE_{\s\n}S^\n_{~\m}+T_{\s\m}\cr
&&\tilde X_{\r\n\m}\equiv X_{\r\n\m}+E_{\r\l}E_{\n\k}Y^{\l\k}_{~~\m}+i E_{\r\l}U^\l_{~\n\m}\cr
&&~~~-iE_{\l\n} V^{~~\l}_{\r\m}-iE_{\r\l\n}S^\l_{~\m}-iE_{\l\n\r}R^\l_{~\m}\label{onstf}
\eea
The known on-shell susy of the usual action specifies that
\bea
&&J_\n^{~\m}+I_\n^{~\m}\equiv \L_\n^{~\m}\cr
&&iM_{\m\n}-E_{\m\r}K^\r_{~\n}=\L_\m^{~\r}E_{\r\n}\cr
&&\tilde T_{\s\m}=-i\L_\s^{~\r}E_{\m\r}\cr
&&2\tilde P_{\k\delta\r}=\tilde P_{[\k\delta]\r}=i\L^{~\n}_{[\k\delta]}E_{\n\r}+i\L^{~\t}_{[\k}E_{\delta]\r\t}\cr
&& \tilde X_{\r\n\m}=-i\L^{~\k}_{\r\n}E_{\m\k}-iE_{\m\n\k}\L^{~\k}_{\r}~, \label{onshell}
\eea
where $\L_\n^{~\m}$ is a complex structure covariantly constant w.r.t. the $+$ connection and preserving the metric $\half E_{(\m\n)}$, and where additional indices denote ordinary derivatives w.r.t. $\P^\m$. When (\ref{onshell}) is satisfied as well as the corresponding relations for the $-$ transformation involving the second complex structure, and we use the $\F$-shell relations we have reduced the theory to that of \cite{Gates:nk}, and we are guaranteed that  the algebra closes on-shell.

From invariance of the action we first obtain
\bea
&&R^\n_{~\s}=iM_{\s\m}E^{\m\n}-J_\s^{~\n}\cr
&&R^\n_{~\s}=E_{\s\t}K^\t_{\m}E^{\m\n}+I_\s^{~\n}\cr
&&(M+T)_{(\m\n)}=0=(M-T)_{[\m\n]\t}\label{Rexp}
\eea
The first two equations are consistent due to (\ref{onshell}). The following two imply (up to a constant) that $T=M$ is antisymmetric, which in turn means that
\beq
K^\r_{~(\m}E_{\n)\r}=\L_{(\m}^{~~\r}E_{\n )\r}=\L_{(\m}^{~~\r}B_{\n )\r}~,
\eeq{Kind}
where we have used (\ref{onshell}) and, in the last equality,  the antisymmetry of $\L_{\m\n}$.
 
Next we have
\bea
&& S^\k_{~\s}=-J_\s^{~\k}-E^{\k\m}\L_\m^{~\r}E_{\r\s}\cr 
&& (I-S)^{(\k}_{~\l}E^{\s ) \l}=0~. \label{Sexp}
\eea
Using the first, the latter relation is satisfied due to a relation that follows from the fact that $\L$ preserves the metric (c.f. (\ref{biherm})). Finally we obtain the following set of conditions:
\bea
&&X_{[\r|\m|\s]}=2P_{\r\s\m}-M_{\m[\r\s]}\cr
&&V_{[\s\r]}^{~~\n}=i\left(M_{[\r|\m}E^{\m\n}\right)_{\s]}+2iP_{\r\s\m}E^{\m\n}\cr
&&Y^{[\s|\n}_{~~\l}E^{\m]\l}=iE^{[\s|\l}I_\l^{~\k|}E^{\m]\n}{~~\k}-2N^{\s\m}_{~~\l}E^{\l\n}\cr
&&iU^\s_{\m\r}+X_{\r\m\t}E^{\s\t}=(M_{\r\l}E^{\s\l})_\m+iK^\s_{[\m\r]}-iQ_{\r\m}^{~\s}\cr
&&V_{\s\l}^{~~\n}E^{\m\l}+iY^{\m\n}_{~~\s}=-J_\s^{~\k}E^{\m\n}_{~~\k}-(K^\m_{~\l}E^{\l\n})_\s
-Q_{\s\l}^{~\m}E^{\l\n}\cr
&&iN^{\r\n}_{~~~\m}+\half U^{[\r}_{~~\m\l}E^{\n]\l}=-\frac{1}{ 4}\left[(I-S)^{[\n}_{~\l}E^{\r]\l}\right]_\m~.\label{acinv}
\eea

To solve (\ref{onshell}) - (\ref{acinv}) we first insert the expressions for $R$ and $S$ from 
(\ref{Rexp}) and (\ref{Sexp}) into the relation for $\tilde T$ in (\ref{onstf}) and (\ref{onshell}). We thus find that $J=\L$ and hence that $I=0$. Next we make use of the symmetries (\ref{fredef}). Requiring that 

\bea
&&\delta\F'_{+\m}(\F',\P)=-\delta\F_{+\m}(\F,\P)+ \delta(2iD_+ \P^\n E_{\n\m})(\F,\P)\cr
&&\delta\F'_{-\m}(\F',\P)=-\delta\F_{-\m}(\F,\P)- \delta(2iD_- \P^\n E_{\m\n})(\F,\P)\label{vfredef}
\eea
tells us that
\beq
Y^{\l\k}_{~~\m}=0, \qquad N^{\l\k}_{~~\m}=0~.
\eeq{NYnoll}

Continuing the analysis, we find that not all coefficient functions in (\ref{tfans}) are determined. This is to be expected, since the general ansatz (\ref{tfans}) will also include ``field equation'' symmetries, i.e., symmetries of the type
\beq
\delta \varphi^i = A_{ij}\frac {\delta {\cal L}} {\delta \varphi^j}~,
\eeq{fesym}
with ${\cal L}$ a Lagrangian for the fields $ \varphi^i$ and $A_{ij}$ some matrix-valued function with the appropriate symmetries.
The set of undetermined functions may be taken to be 
\ber
&P_{\m\n\r}, &\hat K_{\m\n}\equiv E_{[\m |\k}K^\k_{~\n]}\cr
& \hat Q_{\m\n\r}\equiv Q_{(\m |\n|}^{~~~~\k}E_{\r)\k}, &\hat U_{\m\n\r}\equiv E_{(\m |\k}U_{~~\n |\r)}^\k 
\eer{undet}
The $\F_\pm$ field equations  are, from (\ref{fieldeqns}),
\ber
&&F^\m_+\equiv \F_{+\n}E^{\n\m}-i D_+\P^\m\cr
&&F^\m_-\equiv \F_{-\n}E^{\m\n}+i D_-\P^\m~,
\eer{feqdef}
and, for later use , we also define
\ber
&&\tilde F^\m_+\equiv \F_{+\n}E^{\n\m}+i D_+\P^\m\cr
&&\tilde F^\m_-\equiv \F_{-\n}E^{\m\n}-i D_-\P^\m ~,
\eer{tilf}
which on shell become $2i D_+\P^\m$ and $-2i D_-\P^\m$, respectively.

Using the definitions (\ref{undet}) and (\ref{feqdef}), we collect the undetermined functions into an invariance of the type defined in (\ref{fesym}):
\ber
&&\delta \P^\m =0\cr
&&\delta \F_{+\m}=\e^+D_+F^\n_+\hat K_{\n\m}-\e^+F^\l_+D_+\F^\r h_{\r\l\m}\cr
&&\delta \F_{-\m}=\e^+D_+F^\n_-\hat K_{\m\n}+\e^+F^\l_-D_+\F^\r\left(h_{\r\m\l}+\hat K_{\l\m\r}\right)-\e^+D_-\P^\n F^\l_+\hat U_{\l\n\m}~,
\eer{secondinv}
where
\beq
h_{\r\l\m}\equiv iP_{\r\l\m}+\hat Q_{\r\m\l}+\half E_{(\l | \s | \r)}E^{\s\t}\hat K_{\t\m}~.
\eeq{hdef}
Clearly these variations vanish on-shell. It may also be checked that they represent an invariance of the action. In doing this the following relation for the pure $\delta \F$ part of the variation of the Lagrangian is useful:
\beq
\delta_{\F}{\cal L}=\delta \F_{+\m}F^\m_-+F^\m_+\delta \F_{-\m}
\eeq{lvar}

To have the correct on-shell transformations manifest, we rewrite the second supersymmetry using (\ref{feqdef}) and (\ref{tilf}). Removing the transformations (\ref{secondinv}) from (\ref{ansatz}) we are left with \footnote{It is interesting to note that if we relax the condition that $\L_\n^{~\m}$ is covariantly constant w.r.t. the $+$ connection above, we recover it here as a compatibility condition between (\ref{onshell}) and (\ref{acinv}) when $P_{\m\n\r}=0$.}:
\ber
&\delta \P^\m =&\e^+D_+\P^\n \L_\n^{~\m}\cr
&\delta \F_{+\m}=&-\frak 1 2 \e^+D_+F^\n_+E_{\t\n}\L_\m^{~\t}-\frak 1 2\e^+D_+\tilde F^\n_+E_{\t\m}\L_\n^{~\t}\cr
&&-\frak 1 4 \e^+D_+\P^\r F^\l_+\left(-E_{(\l |\s|\r)} E^{\s\t}\L_{(\m}^{~~\k}B_{\t )\k}+4i\tilde P_{\r\l\m}\right)+\e^+D_{+}\P^\r D_{+}\P^\n \tilde P_{\r\n\m}\cr
&\delta \F_{-\m}=&\e^+D_+F^\n_-R^\l_{~\m}E_{\l\n}+\e^+D_-F^\n_+(E_{\n\l}S^\l_{~\m}-\frak 1 2 \L_\n^{~\r}E_{\m\r})+\frak 1 2\e^+D_-\tilde F^\n_+\L_\n^{~\r}E_{\m\r}\cr
&&-\e^+D_+\P^\r F^\n_-\left( \frak 1 2 E_{(\m |\s|\r)} E^{\s\t}\L_{(\n}^{~~\k}B_{\t )\k}+i\tilde P_{\m\r\n}+E_{\l\n}E_{\m\t}\L_\r^{~\k}E^{\t\l}_{~~\k}+\frak 1 2 (\L_{(\m}^{~~\t} E_{\t|\n)})_\r\right)\cr
&&+\frak 1 2 \e^+F^\r_+D_-\P^\n(S^\k_{~~\r\n}E_{\m\k}-E_{\r\l\n}S^\l_{~\m})+\e^+D_{+}\P^\r D_{-}\P^\n \tilde X_{\r\n\m}~,
\eer{psec}
where now
\ber
&&S^\k_{~\m}=E^{\k\t}\L_{[\t}^{~~\l}B_{\m]\l}\cr
&&R^\k_{~\m}=\frac 1 2 E^{\t\k}\L_{(\t}^{~~\l}B_{\m)\l}~.
\eer{news}
Although the transformations no-longer contain any undetermined functions, there are still some ambiguities left.
First we may shift the coefficients in front of $F_\pm$ and $\tilde F_\pm$ using 
\beq
F^\m_\pm -\tilde F^\m_\pm= \mp 2i D_\pm\P^\m~.
\eeq{frel}
Second, we may use $\nabla^{(+)}\L=0$ to change the terms containing derivatives of the complex structure, e.g, and finally we may still identify and remove additional ``field equation'' symmetries. In fact, using the first and last option we find our final form of the variations

\ber
&\delta \P^\m =&\e^+D_+\P^\n \L_\n^{~\m}\cr
&\delta \F_{+\m}=&-\frak 1 2 \e^+D_+F^\n_+E_{\t\n}\L_\m^{~\t}-\frak 1 2\e^+D_+\tilde F^\n_+E_{\t\m}\L_\n^{~\t}+\e^+D_{+}\P^\r D_{+}\P^\n \tilde P_{\r\n\m}\cr
&\delta \F_{-\m}=&\frak 1 2 \e^+D_+F^\n_-E_{\n\t}\L_\m^{~\t}+\frak 1 2\e^+D_+\tilde F^\n_-\L_\n^{~\t}E_{\m\t}+\e^+D_{+}\P^\r D_{-}\P^\n \tilde X_{\r\n\m}\cr
&&+\e^+D_+\P^\r F^\n_-\left( E_{\m\n\k}\L_\r^{~\k}-\frak 1 2 (\L_{(\m}^{~~\t} B_{\t|\n)})_\r\right)~.
\eer{ppsec}
Again, an explicit check confirms the invariance of the action.

\Section{Discussion}

The previous derivation was focused completely on the $\e^+$-symmetry. In fact, the $\e^-$-symmetry follows trivially from that discussion. All we have to do is make the exchange 
\ber
&&+ \leftrightarrow -\cr
&&E^{\m\n}\leftrightarrow -E^{\n\m}\cr
&&\L\equiv \L^{(+)}\leftrightarrow  \L^{(-)}~,
\eer{minus}
in (\ref{ppsec}). The on-shell invariance under the combined variations is then guaranteed by the  $\F$-shell equivalence to the usual model.

In section 3 we mention that the model (\ref{betteraction}) that we study becomes a supersymmetrization of the  Poisson sigma model when we set the metric $G_{\m\n}$ to zero. Unfortunately, our entire treatment above rests on having a non-zero hermitean metric, and we are thus at present unable to draw any conclusions about the existence of a  second supersymmetry in this case. The opposite limit of a zero 
$B_{\m\n}$ field is readily treated, however. The transformations (\ref{ppsec}) then reduce to
\ber
&\delta \P^\m =&\e^+D_+\P^\n \L_\n^{~\m}\cr
&\delta \F_{+\m}=&\e^+\partial_\+\P^\n\L_{\n\m}+i\e^+D_{+}\P^\r D_{+}\P^\n\tilde X_{\r\n\m}\cr
&\delta \F_{-\m}=&iD_+D_-\P^\n\L_{\n\m}-i\e^+D_{+}\P^\r D_{-}\P^\n \tilde X_{\r\n\m}\cr
&~~~~~~~~~~~~~~&+\e^+D_+\P^\r F^\n_-G_{\m\n\k}\L_\r^{~\k}~,
\eer{gnull}
where now
\beq
\tilde X_{\r\n\m}=\left(\L_{\r\m\n}+\L_\r^{~~\t}G_{\m[\n\t]}\right)~.
\eeq{}
The only $\F$ dependence is thus through $F_-$ in the last term in (\ref{gnull}).
\Section{Isometries, gauging and duality}

Under an isometry of the target space
\ber
&&\delta\P^\m=\e k^\m(\P)\cr
&&\delta\F_{\pm\m}= -\e\F_{\pm\n}k^\n_{~\m}(\P)\cr
&&{\cal L}_{\e k}E^{\m\n}=0~,
\eer{isom}
the action (\ref{betteraction}) stays invariant. Such an isometry is promoted to a local invariance ($\e\to\e(\xi,\theta)$) by the substitution (for more general cases see \cite{Hull:1985pq})
\beq
D_\pm \P^m\to D_\pm \P^m+A_\pm k^\m~,
\eeq{covder}
where  $\delta A_\pm=-D_\pm\e$. We now use this local invariance to discuss duality\footnote{For introductions to duality see e.g., \cite{Giveon:1994mw}-\cite{Bakas:1995hc}. Some early discussions in superspace are \cite{Lindstrom:rt,Howe:sb}}.

The gauged action reads, including a Lagrange multiplier $Y$  ensuring that $A$ is pure gauge:
\beq
    S= -\int d^2\xi d^2\theta\,\, \left\{\F_{+\m} \F_{-\n} E^{\mu\nu}(\Phi)+i\F_{(+\m}\nabla_{-)} \Phi^\m-A_{(+}D_{-)}Y \right\}~,
\eeq{gaugedbetteraction}
where
\beq
\nabla_\pm \P^\m\equiv D_{\pm} \Phi^\m+A_{\pm}k^\m
\eeq{}
Integrating out $Y$ and choosing gauge, we recover the original action (\ref{goodaction}). If we integrate out the gauge field $A_\pm$ instead, we find 
\beq
D_\pm Y = i\F_{\pm\m}k^\m~.
\eeq{}
In adapted coordinates wher $k^\m = \delta^{\m 0}$ and in a gauge where $D_\pm\P^0=0$, the dual action then reads
\beq
    S=- \int d^2\xi d^2\theta\,\, \left\{\F_{+i} \F_{-j} E^{ij}+D_{+}Y \F_{-i} E^{0i}+\F_{+i}D_{-}Y  E^{i0}+D_{+}YD_{-}Y  E^{00}+i\F_{(+i}D_{-)} \Phi^i \right\}~,
\eeq{dualbetteraction}
where $i,j=1...d-1$. To recognize the ``second order'' form of this action we integrate out $\F_{\pm\m}$ which yields the equations
\ber
&&\F_{-j}E^{ij}+D_-YE^{i0}+iD_-\P^i=0\cr
&&\F_{+j}E^{ji}+D_+YE^{0i}-iD_+\P^i=0~.
\eer{}
Solving these equations and substitutiong into (\ref{dualbetteraction}) we obtain 
\beq
    S= \int d^2\xi d^2\theta\,\, D_{+} \Phi^\mu D_{-} \Phi^\nu \tilde E_{\mu\nu}
    (\Phi),
    \eeq{dualgoodaction}
where
\ber
&&\tilde E_{00}= E^{-1}_{00}\cr
&&\tilde  E_{0i}= E^{-1}_{00}E_{0i}\cr
&&\tilde  E_{i0}= E^{-1}_{00}E_{i0}\cr
&&\tilde  E_{ik}= (E_{ik}-E^{-1}_{00}E_{i0}E_{0k})~,
\eer{}
i.e., the usual Busher rules.\cite{Buscher:sk,Buscher:qj,Hassan:mq}
\Section{Conclusions}
In this note we have addressed the question of whether a formulation of the ${\cal N}=(2,2)$ supersymmetric non-linear sigma model with auxiliary superfields in $^ *{\cal T}$ has a richer target space geometry than the original one. Under the assumptions made in the paper, which seem to be quite general, we find that the second supersymmetry is determined by the same comlex structures that determine the transformations in the original model plus the background metric and $B$-field. It would seem natural to assume that this indicates a relation to the generalized complex geometry of the type discussed in \cite{nigel,marco}, since that geometry is also determined by these objects (although the $B$-field there is closed). This is a topic for further study, however.

As mentioned in  section 5, an open problem is to find the second supersymmetry when the metric is zero and we have a pure Poisson sigma model.

Another interesting question is what happens if we require less supersymmetry, ${\cal N}=(2,1)$ say. Then  our action may be extended by inclusion of kinetic terms for the $\F$-fields since those are no-longer auxiliary. The geometry of such models needs to be investigated.

Finally, a recent investigations on the boundary conditions for open models 
\cite{Albertsson:2001dv}- \cite{Melnikov:2003zv} generalize in interesting ways to the present model.

\bigskip

{\bf Acknowledgements:} I thank Martin Ro\v cek for comments and  construcive critisism, and dedicate this paper to him on the occation of his 50th birthday. Discussions with Fiorenzo Bastianell, Paul Howe, Maxim Zabzine and Konstantin Zarembo are gratefully acknowledged as well as the hospitality and stimulating atmosphere of the  Simons Work Shop at the C.N. Yang Institute, Stony Brook and the Department of Physics at the University of Bologna. This work was supported in part by VR grant 650-1998368.

 \end{document}